\documentclass[twocolumn,10pt,pra]{revtex4-1}

\def\by#1{#1,}
\def\and{and }
\def\yr#1{{(#1)}}
\def\paper#1{#1}
\def\jour#1{{\it #1}}

\def\vol#1{{\bf #1},}
\def\issue#1{}
\def\pages#1{\hbox{#1},}

\usepackage{amsmath}
\usepackage{amsfonts}
\usepackage{amssymb}
\usepackage{graphicx}
\usepackage{appendix}
\usepackage{placeins}
\usepackage{multirow}

\begin{document}

\title{Capillary transport in paper porous materials at low saturation levels.}

\author{Alex V. Lukyanov}
\email{corresponding author,\\ a.lukyanov@reading.ac.uk}
\affiliation{School of Mathematical and Physical Sciences, University of Reading, Reading, RG6 6AX, UK}
\affiliation{P.N. Lebedev Physical Institute of RAS, 119991, Moscow, Russia}

\author{Vladimir V. Mitkin}
\affiliation{Aerospace Research Laboratory, University of Virginia, Charlottesville, VA 22903, USA}

\author{Tristan Pryer}
\affiliation{Department of Mathematical Sciences, University of Bath, Bath, BA2 7AY, UK}

\author{Penpark Sirimark}
\affiliation{Department of Science and Mathematics, Rajamangala University of Technology Isan, Surin, Thailand}

\author{Theo G. Theofanous}
\affiliation{University of California, Santa Barbara, CA 93106, USA}

\begin{abstract}
The problem of capillary transport in fibrous porous materials at low levels of liquid saturation has been addressed. It has been demonstrated, that the process of liquid spreading in this type of porous materials at low saturation can be described macroscopically by a similar super-fast, non-linear diffusion model as that, which had been previously identified in experiments and simulations in particulate porous media. The macroscopic diffusion model has been underpinned by simulations using a microscopic network model. The theoretical results have been qualitatively compared with available experimental observations within the witness card technique using persistent liquids.           
\end{abstract}

\maketitle
 
\section{Introduction}
Liquid distributions and transport in particulate porous media, such as clay, loam and sand, at low saturation levels was found to have very distinctive features resulting in a special class of mathematical problems, {\bf superfast non-linear diffusion}~\cite{Lukyanov2012, Penpark2018, Penpark2019, Lukyanov2019}. It has been established, both experimentally  and theoretically, that any time any wetting liquid naturally (that is when there is no force wetting regime involved) spreads in a dry (or a nearly dry) porous particulate matrix, the moving front dynamics follows, after some time, the power evolution law dictated by the super-fast non-linear diffusion mechanisms.  

The {\bf special character} of this non-linear diffusion process is caused by the loss of global, pore-scale connectivity at low levels of saturation. In this case the liquid transport only occurs over the surface elements of the porous matrix, sand particles for example, while the liquid is mostly located in the capillary bridges formed at the point of particle contacts. 

Apparently, liquid spreading in dry porous materials is not just characteristic for particulate porous media, such as sand, but also frequently occurs in other porous materials. Therefore, in this paper, the previously developed model is generalised to another fairly common type of porous materials consisting of fibre elements, such as papers and textiles, where a similar kind of non-linear diffusion process is anticipated. We would like to understand how general and universal the super-fast mechanism, first discovered in particulate porous media~\cite{Lukyanov2012}, actually is. We note that this transport regime is important for many medical applications, since it commonly occurs in {\bf chromatographic flows and lateral flow test setups} widely used in the infectious disease testing~\cite{Lateral-flow-review, Paper-micro-fluidic-review-2017, Paper-micro-fluidic-2020}.     

\section{Macroscopic and microscopic models of capillary transport in fiber porous materials}

The structure of fibrous porous materials is quite different from that of particulate porous media~\cite{Fibre-materials-1, Fibre-materials-2, Fibre-materials-3, Fibre-materials-31, Fibre-materials-4, Herminghaus-2005}. Yet, all the main elements of the super-fast diffusion model can be identified here too. 

At low saturation levels, the liquid is only located on the surfaces of the fibres (including intrafibre pores) and in the liquid bridges formed at the intersections of the fibres~\cite{Fibres-Bridge-1, Fibres-Bridge-2, Fibres-Bridge-3}. The microscopic surface details, such as roughness, generate the capillary pressure to drive the liquid flow through the network, where the liquid bridges, as in the case of particulate media, play the role of variable volume reservoirs. We further assume that the liquid at least partially wets the fibres, so that the contact angle on the rough surfaces of the fibres would be small (close to zero) or zero. 

As in the previous case of particulate porous media~\cite{Lukyanov2019}, we first consider the morphology of the liquid distribution in fibrous materials at low saturation levels to formulate a pressure-saturation relationship, which will be further used to obtain a macroscopic, average model. The macroscopic model will be compared with a microscopic network setup. In the end, we consider a set of available experimental data, and compare the general trends expected from the macroscopic formulation with the experimental results.    

\subsection{Quasi-steady liquid distribution in fibrous materials at low saturation levels}

The morphology of the liquid structures formed between the crossing fibres in the wetting case is found to be in general more complex than that observed between the particles~\cite{Herminghaus-2005, Fibres-Bridge-1, Fibres-Bridge-2, Fibres-Bridge-3, Herminghaus-2008, Herminghaus-2008-2}. 

In particulate porous media, isolated bridges only exist below a certain critical level of saturation $s\le s_c\approx  8-10\%$, where the saturation $s$ is defined as the ratio of the liquid volume $V_L$ to the available volume of voids $V_E$, $\displaystyle s=\frac{V_L}{V_E}$ in a sample volume $V$. Above the critical level, isolated bridges coalescence into larger clusters, such as trimmers, pentamers and more complex agglomerates. This trend has been observed for idealized systems consisting of spherical grains and for non-spherical particle media, like real sand~\cite{Herminghaus-2008}. 

In fibrous porous media, the liquid volume at the crossing of two rigid fibres can take under the action of surface tension forces several distinct morphologies depending on the amount of the liquid $V_B$, the separation distance and the angle between the fibres $\theta_{f}$, Fig. \ref{Fig1}: a long liquid column, a mixed morphology state that consists of a drop on one side together with a small amount of liquid on the other side and a drop or a compact hemispherical drop or a pendular ring~\cite{Fibres-Bridge-1, Fibres-Bridge-2, Fibres-Bridge-3}. In general, the elongated liquid columns are only formed at small angles $\theta_f\le 20^{\circ}$ between the crossing fibres~\cite{Fibres-Bridge-2}. So the predominant shape of the liquid volumes in randomly oriented fibrous materials appears to be either a drop or a pendular ring at small volumes $V_B\ll V_E$.

The shape of an isolated bridge (a pendular ring) can be determined analytically in a closed form only in quasi-static conditions and in a simplified geometry, for example in the case of two spheres in contact or at small separating distances~\cite{Orr-Scriven-1975}. The analytic forms are quite lengthy, but approximately results in the following scaling of the capillary pressure in the liquid bridge $p$ as a function of its volume $V_B$ 
\begin{equation}
\label{Paper-pressure-law}
p\approx -p_0 \left(\frac{R^3}{V_B} \right)^{\gamma_p},\quad \gamma_p\approx1/2.
\end{equation}
Here, the length scaling parameter $R$ could be either the diameter of the spherical particle (in particulate porous media) or the characteristic fibre thickness, $p_0=\frac{4\gamma}{R}$ and $\gamma$ is the coefficient of surface tension of the liquid~\cite{Halsey1998, Herminghaus-2008, Lukyanov2012, Lukyanov2019}. The scaling law can be applied at low levels of saturation $s\le s_c$, even for particulate media consisting of non-spherical particles, such as sand, before the capillary pressure saturates at a universal critical value~\cite{Herminghaus-2008}. As a result, in what follows, relationship (\ref{Paper-pressure-law}) is taken as the main pressure scaling law in capillary bridges at fibre crossings in our model. The scaling law is expected to be violated only if the dominant morphology of the liquid volumes would change from a drop (or a pendular ring) to elongated columns.  

\subsection{Macroscopic and microscopic parameters of fibrous materials}

We note that the connectivity of fibres in a porous material (the main morphology of the crossings) can be also in a form of a branch, when each crossing has three links coming out instead of four as in the case of normal crossing, Fig. \ref{Fig1}.  

The porosity of fibrous materials $\phi$ is highly variable (one can easily change paper porosity by applying moderate mechanical pressure to a sample), and, in general, it is much higher than that of particulate porous media. The typical porosity values for most paper grades are found to be around $\phi=0.7$ (sand porosity, in comparison, is around $\phi=0.3$)~\cite{Fibre-materials-4}. The larger porosity values imply that overlapping (coalescence) of the liquid volumes attached to different crossings (the effect observed in particulate porous media~\cite{Herminghaus-2005, Herminghaus-2008}) may only occur at much larger values of saturation.

It is well known that the structure of fibrous materials is effectively two-dimensional, that is the fibres are roughly oriented in the paper sheet plane. The main characteristics of the paper materials are therefore also two-dimensional, such as the total length of fibres $L_q$ per unit area of a paper sheet.  Typically, it takes the values in between $200 \le L_q \le 400 \, \mbox{mm}^{-1}$ at the characteristic paper thickness around $50\,\mu\mbox{m}$~\cite{Fibre-materials-31}. Given the characteristic fibre thickness $R$ in the range $4\,\mu\mbox{m} \le R \le 10\,\mu\mbox{m}$, one can define the total length of fibres $L_e$ per unit area in a layer of thickness $R$, which is expected in the range of $16\le L_e \le 80 \,\mbox{mm}^{-1}$. The so-obtained typical range is consistent with the typical paper porosity levels. Indeed, $\phi=\frac{V_E}{V}$, that is $\phi = 1-\pi L_e R/4$ in a sample volume $V$ of thickness $R$ assuming circular fibre cross-section area $\pi R^2/4$. The estimate then gives $\phi\approx 0.7$, if we take parameters in the middle of their expected, estimated intervals, that is $L_e = 50 \,\mbox{mm}^{-1}$ and $R = 7\,\mu\mbox{m}$. This implies that parameters $\phi$, $L_e$ and $R$ characterising porous network are always interrelated.       

\subsection{Macroscopic model}

To characterise liquid distributions macroscopically, one needs to introduce quantities averaged over a sufficiently large volume element. In what follows, we will briefly follow the procedure similar to that in~\cite{Lukyanov2019}, while binding parameters to the specific case of fibrous materials and defining their characteristic values. 

First, an average coordination number, that is the average number of crossings per unit volume $N_c$ is to be defined. The value of $L_e$ in a random paper network allows to estimate the mean distance $l_f$ between the nearest fibre crossings, as in~\cite{Fibre-materials-31}, $l_f\approx \frac{2}{\pi L_e}$. That is, typical values of $l_f$ are expected in the range $8\,\mu\mbox{m}\le  l_f\le 13\,\mu\mbox{m}$. Using $R = 7\,\mu\mbox{m}$ and $L_e = 50 \,\mbox{mm}^{-1}$, one can obtain an estimate of the coordination number with a typical value $N_c = \frac{\pi L_e^2}{2 R} \approx 5.6\times 10^5 \,\mbox{mm}^{-3}$. 

To parametrize saturation, we split, similar to~\cite{Lukyanov2019}, average liquid content in a sample volume $V=S_0 R$ of thickness $R$ and surface area $S_0$ into two parts: the liquid contained on the rough surface of fibres and in the intrafibre pores of volume $V_r= L_e S_0 \delta_R^2$ and the liquid contained in the capillary bridges at the fibre crossings $V_c=V_B N_c\, V$. The parameter $\delta_R$ has the dimension of length and can be interpreted as the characteristic length scale of the surface roughness (intrafibre pore size), which could be considered as the fitting parameter of the model. We further assume that the smaller details (on the length scale $L\ll R$) of fibres are fully saturated, as it is commonly found on the rough surfaces~\cite{Yost-1998}, such that the amount of the liquid stored  on the rough surface of fibres and in the intrafibre pores is independent of the liquid pressure, that is constant.
This approximation is well fulfilled if the capillary pressure is on the scale of $p\approx p_0$.

Combining both contributions, saturation 
$$
s=\frac{V_c+V_r}{\phi V}
$$ 
can be presented as
\begin{equation}
\label{saturation}
s=V_B V_0^{-1} + s_0, \quad V_0=\frac{\phi}{N_c},
\end{equation}
where
$$
s_0=\frac{L_e \delta_R^2}{\phi \, R}
$$
is the saturation level when all liquid bridges cease to exist.

Then, using (\ref{saturation}), the average capillary bridge pressure $P=<p>^l$ 
\begin{equation}
\label{Pressure-saturation}
P=-p_0\left( \frac{R^3}{V_0}\right)^{1/2}\frac{1}{(s-s_0)^{1/2}},
\end{equation}
where
$<...>^l=V_l^{-1}\int_{V_l} d^3 x$ is intrinsic liquid averaging, $V_l$ is liquid volume within the sample volume $V$. Using $L_e = 50 \,\mbox{mm}^{-1}$, $\delta_R=1\,\mu\mbox{m}$, $R = 7\,\mu\mbox{m}$ and $\phi=0.7$ as the typical parameters, one can estimate that the residual saturation level $s_0\approx 10^{-2}$, that is about $1\%$ as expected. 

Consider now local transport on the surface of fibres and in the intrafibre pores. The surface flux density ${\bf q}$, according to the previous study of liquid spreading on rough surfaces made of microscopic grooves of various shapes and dimensions~\cite{Yost-1998}, obeys a Darcy-like law
\begin{equation}
\label{Micro-Darcy-1}
{\bf q}=-\frac{\kappa_m}{\mu} \nabla \psi,
\end{equation} 
where $\mu$ is liquid viscosity, $\psi$ is local pressure in the liquid averaged within the surface roughness and $\kappa_m=\kappa_0 \delta_R^2$ is the effective coefficient of permeability of the surface roughness, which is proportional to the square of the length scale parameter $\delta_R$. In the assumption of fully saturated fibres, $\kappa_m=const$.

According to the spatial averaging theorem~\cite{Whitaker-1969}, applying intrinsic liquid averaging $<...>^l$
\begin{equation}
\label{Macroscopic-Darcy-1}
-\frac{\kappa_m}{\mu} \left\{ \nabla <\psi>^l + V_l^{-1}\, \int_{S_l} \psi\, {\bf n}\, dS \right\}=<{\bf q}>^l, 
\end{equation} 
where $S_l$ is the area of liquid interface with normal vector $\bf n$. The surface integral in the creeping flow conditions, when the pressure variations across the liquid layer are insignificant, can be neglected $V_l^{-1}\, \int_{S_l}\, \psi\, {\bf n}\, dS\approx 0$ and 
\begin{equation}
\label{Macroscopic-Darcy-2}
-\frac{\kappa_m}{\mu}  \nabla <\psi>^l =<{\bf q}>^l. 
\end{equation} 
Now, one can cast the continuity equation, in the absence of evaporation,
$$
\frac{\partial (\phi s)}{\partial t} + \nabla\cdot {\bf Q}=0
$$
into
\begin{equation}
\label{Gov-1}
\frac{\partial (\phi s)}{\partial t}=\nabla\cdot \left\{ \frac{K}{\mu}  \nabla P\right\}.
\end{equation}
Here, 
\begin{equation}
\label{Def-SE}
{\bf Q}=\frac{S_e}{S}<{\bf q}>^l,
\end{equation} 
$S$ is the surface area of the sample volume $V$ with the effective area of entrances and exits $S_e$ and coefficient $K=\kappa_m \frac{S_e}{S}$. Also, it has been assumed that in the creeping flow conditions $P= <p>^l \approx <\psi>^l$. Note, that the ratio $S_e/S$ is not strictly speaking just a geometric factor. It is an average quantity defined by (\ref{Def-SE}), which incorporates connectivity and the shape of the surface elements. 

Assuming further that porosity $\phi$ is constant and using expression (\ref{Pressure-saturation}) for the average pressure, one can transform the governing equation (\ref{Gov-1}) into a non-linear diffusion equation for the saturation $s({\bf x},t)$
\begin{equation}
\label{Superfast-1}
\frac{\partial s}{\partial t}= \nabla\cdot  \left\{  \frac{D_0 \, \nabla s}{(s-s_0)^{3/2}}\right\}, 
\end{equation}
where
$$
D_0=\frac{1}{2}\frac{K}{\mu}\frac{p_0 }{\phi }\left( \frac{R^3}{V_0}\right)^{1/2}. 
$$

The resultant non-linear diffusion equation (\ref{Superfast-1}) has a similar form as that studied in~\cite{Lukyanov2012, Lukyanov2019} in the case of particulate porous media. The main difference at this point is that the equation in the bulk has a constant coefficient of diffusion $D_0$, which is defined by the connectivity of the porous network of fibres, while in particulate porous media, there is a weak logarithmic dependence on saturation, and the diffusivity is driven by the shape of the particles and their contact area, details can be found~\cite{Lukyanov2012, Penpark2018, Penpark2019, Lukyanov2019}. In a way, the situation is simpler in the case of fibrous materials than that in particulate media, since the connectivity parameter can be quite accurately found via a network model. This will be done in the next part of this study. On the other hand, the question of the liquid amount stored in the intrafibre space is still open, and down to simplifying assumptions at this stage. 

To address a moving boundary value problem set in an open domain with a smooth boundary $\partial \Omega$ moving with velocity ${\bf v}$, the governing equation (\ref{Superfast-1}) can be complemented with the boundary conditions
\begin{equation}
\label{BC1}
\left. s\right|_{\partial \Omega}=s_f, \quad s_f>s_0
\end{equation}
and
\begin{equation}
\label{BC2}
{\bf v\cdot n}=-D_0\frac{{\bf n}\cdot\nabla s}{s(s-s_0)^{3/2}},
\end{equation}
where $\bf n$ is the normal vector to the boundary $\partial \Omega$. 

\subsection{The boundary value of saturation and steady states}

The existence of a sharp boundary during the wetting of a dry porous material has been established experimentally in the case of particulate porous media~\cite{Lukyanov2012, Lukyanov2019}, in the experiments with the paper porous materials, a sharp boundary was also observed, though there are some differences discussed below. 

As we have shown previously, the boundary value of saturation $s_f$ is defined by the capillary pressure developed at the moving front, which in turn is conditioned in particulate porous media by the formation of bottleneck regions at the point of particle contacts~\cite{Lukyanov2019}. In the fibrous porous media, such clear separation of the length scales generating the capillary pressure is not expected in a general case. Indeed, while the contact area between two particles vanishes when the bridge size shrinks and bottleneck regime of the contacts is achieved, the contact area between the fibres is expected to be still of the order of the fibre diameter $R$. In particulate porous media, this leads to a sharp cut off when propagation of the moving front practically stops. In the fibrous materials, this transition should be smoother to the mode, when the transport will be mostly conditioned by the smaller details of the fibres, for example intrafibre pores or other smaller elements of a fibre. Further in the model development, we consider only the regime when the liquid bridges still exist, so that the minimal level $s_0$ is defined by fully saturated intrafibre structure. The boundary value then is always supposed to be larger then the minimal value $s_f>s_0$ and should be defined by the length scale of the fibre details, $\delta_R$.

To get an estimate of the typical values of the boundary pressure and the saturation, we assume that the pressure is generated by the capillaries with characteristic size of the order of $\delta_R$. Then, for example for water, taking characteristic value of the surface tension $\gamma=72\,\mbox{mN}/\mbox{m}$ at $25^{\circ}\, \mbox{C}$, one can obtain that at $\delta_R=1\,\mu\mbox{m}$ the capillary pressure $\displaystyle P=\frac{2\gamma}{\delta_R}\approx 1.4\times 10^{5}\,\mbox{Pa}$. As a result, from (\ref{Pressure-saturation}), taking typical parameter values $L_e = 50 \,\mbox{mm}^{-1}$, $R = 7\,\mu\mbox{m}$ and $\phi=0.7$, parameter $s_f=0.022$, that is $s_f\approx 2.2 \, s_0$ at similar values of parameters. 

One needs to note though, that in general the capillary pressure at the moving front may be generated by the fibre irregularities of smaller length scale than the average typical values responsible for the liquid accumulation in the fibres, that is contributing into the value of parameter $s_0$. So that parameters $s_f$ and $s_0$ strictly speaking can be regarded as independent. 

\subsection{Microscopic model}

As it follows from the macroscopic formulation (\ref{Superfast-1}), to accurately predict liquid spreading at low saturation levels, one needs to know the main parameter $K=\kappa_m\frac{S_e}{S}$ contributing into the diffusivity, which, in turn, is defined by the connectivity of the porous paper network, that is by the parameter $S_e/S$. Connectivity is essentially a microscopic quantity, which can be only obtained using a microscopic network model. 

Another reason to turn to a microscopic view, that is to a network model here is to underpin the macroscopic formulation and, what's more important,  to establish sensitivity of the connectivity factor $S_e/S$ to the conducting properties of the fibres and their distribution. While modeling the transport in porous media using network models has shown, in general, that the methodology is stable and reliable, and is able to converge to the macroscopic results ~\cite{Meyers-1999, Blunt-2001, Sousa-2009}, the super-fast diffusion has anomalous properties, such as a divergent coefficient of diffusion, so that this would be interesting and informative to verify the macroscopic formulation in this case.  

\begin{figure}[ht!]
\begin{center}
\includegraphics[trim=0.5cm 2.cm 0.5cm 0cm,width=0.8\columnwidth]{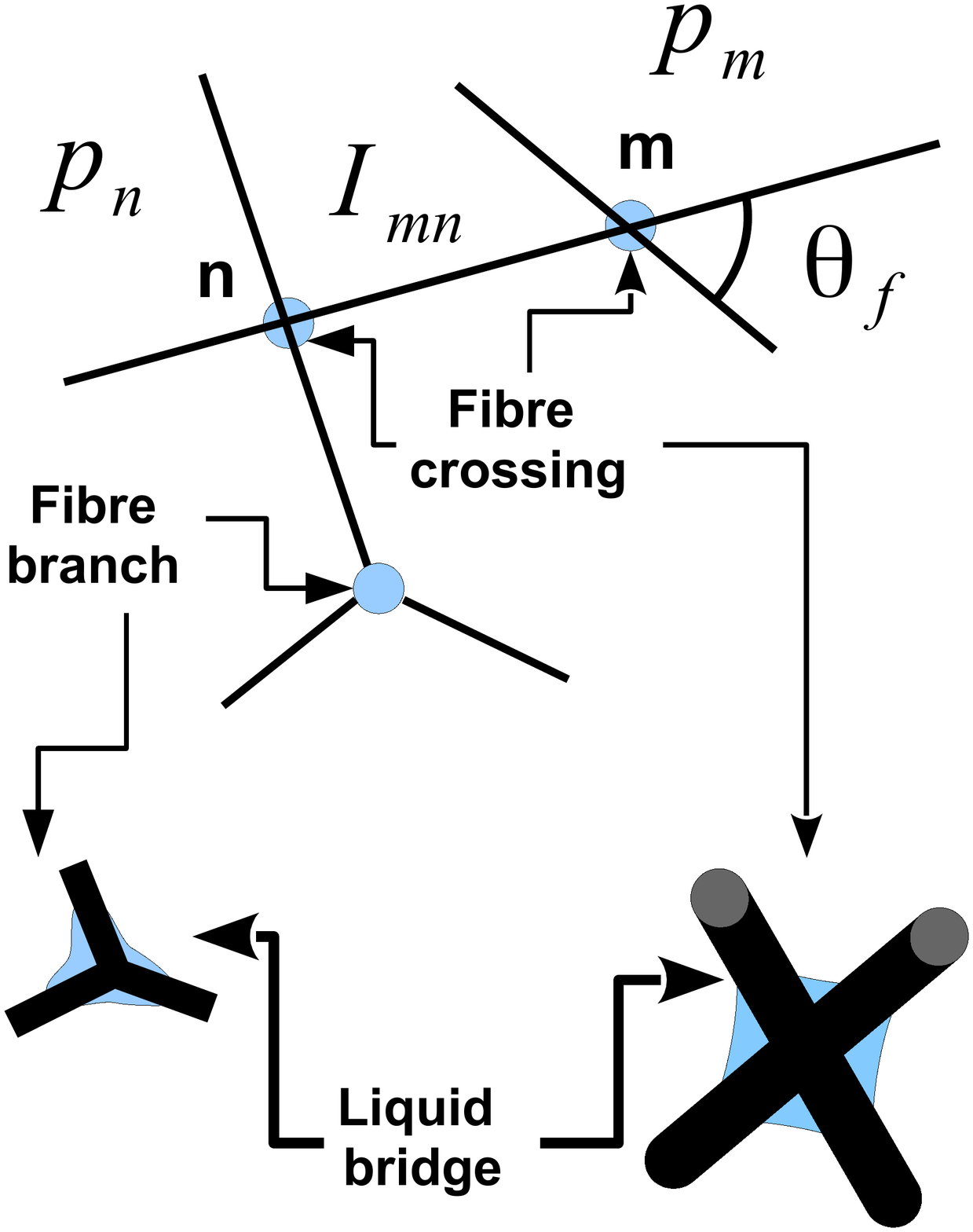}
\end{center}
\caption{Illustration of the fibre crossing and the microscopic network.} 
\label{Fig1}
\end{figure}

The microscopic network model, we use here, is based on some simplifying assumptions. First of all, the microscopic network is essentially two-dimensional and consists of two elements: randomly placed nodes corresponding to the liquid bridges at the paper fibre crossings or at the branch points and the links corresponding to the fibres connecting the bridges, see Fig. \ref{Fig1}. The random distribution of nodes has been generated using Voronoi algorithm and Delaunay triangulations~\cite{Fortune-1995}, when the original domain of simulations is tessellated into either triangles (three neighbours per node) or quadrilaterals (four neighbours per node), Fig. \ref{Fig1}.   

To obtain an equivalent to a three-dimensional case distribution of nodes in the two-dimensional network, the total number of nodes per unit area $N_s$ in the two-dimensional case is set to be the same as the total number of crossings per unit area in a porous layer of thickness $R$, that is $N_s=N_c R$. 

In what follows, we will use non-dimensional forms by normalizing distance, pressure, flux density and flux by $R$, $p_0$, $q_0=\frac{\delta_R}{R}\frac{k_m p_0}{\mu R}$ and $I_0=\pi \delta_R \frac{k_m p_0}{\mu}$ respectively. Then in non-dimensional form, designating non-dimensional variables by a bar, $\bar{N}_s=\bar{N}_c=\frac{8}{\pi}(1-\phi)^2$. 

We further assume, according to (\ref{Paper-pressure-law}), that at any node the liquid pressure is defined by the amount of the liquid in the bridge, that is
$$
\bar{p}_m=-\left(\frac{1}{\bar{V}_B^m} \right)^{1/2},
$$ 
where $\displaystyle \bar{V}_B^m$ is the normalized liquid volume at node $m$. At the same time, at non-equilibrium, the liquid flux between the nodes connected through the links is defined by the pressure difference. That is the liquid flux from node $m$ to node $n$ connected by the filament of length $\bar{L}_{mn}$ is proportional to the pressure difference between the nodes 
$$
\bar{I}_{mn}=-  \alpha_{mn}\frac{\bar{p}_n-\bar{p}_m}{\bar{L}_{mn}}.
$$
The coefficient of proportionality here is a non-dimensional adjusting parameter $\alpha_{mn}$, which takes into account the average shape of the fibres and their average ability to transport the liquid. In our simulations, parameter $\alpha_{mn}$ is either a constant, $\alpha_{mn}=1$, or is a random number uniformly distributed in the interval $0\le \alpha_{mn} \le 1$, such that the average $<\alpha_{mn}>=1/2$. Note, if all links would be of a cylindrical shape of diameter $\bar{D}_R=1$ having a uniform liquid layer of thickness $\delta_R/R$ carrying the liquid flux, then all $\alpha_{mn}=1$.

Liquid saturation can then be calculated as an average over some representative (that is containing many nodes) surface element with surface area $\bar{S}_0$. Since our prime concern here is permeability of the network, we will neglect the amount of the liquid stored in the links. That is, 
$$
s=\frac{\sum_k \, \bar{V}_B^k}{\phi \bar{S}_0 },
$$ 
where the summation is over all nodes within the surface element. 

In a non-equilibrium state, the distribution of liquid in the network evolves in time with a time step $\Delta t$ chosen to achieve numerical stability. After each time step, the total amount of the liquid at every node is calculated according to the mass balance, that is the mass change due to the total flux through the links connected to the node and the amount at the previous time step. 

Our prime concern here is a steady state when the flux density is constant. To obtain informative data, we setup a quasi one-dimensional problem. In the setup, the two-dimensional square area (side size $\bar{X}=100$) is divided into equidistant strips in the $x$-direction (the direction of the diffusion) of a fixed width $\Delta_x=2.5$. The nodes in the first and in the last strip are kept at a fixed liquid volume to emulate fixed boundary saturation levels. In the perpendicular to the $x$-direction, the $y$-direction, zero flux boundary condition is set. The setup is supposed to be equivalent to a one dimensional problem for  (\ref{Superfast-1}) with Dirichlet type boundary conditions, that is in a steady state
$$
 \frac{\partial }{\partial \bar{x}} \left\{ \frac{\partial s}{\partial \bar{x}} s^{-3/2}\right\}=0,\quad x\in(0,\bar{X})
$$ 
$$
s(0)=s_1>0, \quad s(\bar{X})=s_2>0.
$$

The differential equation has a general solution 
\begin{equation}
\label{S-profile}
s=\frac{1}{(\bar{C}_0 + \bar{C}_1 \bar{x})^2}, 
\end{equation}
where 
$$
\quad \bar{C}_0=\frac{1}{\sqrt{s_1}}, \quad \bar{C}_1=\frac{1}{\bar{X}}\left(\frac{1}{\sqrt{s_2} } - \frac{1}{\sqrt{s_1}}\right).
$$

The constant flux density then
\begin{equation}
\label{qs}
\bar{q}_s=-\frac{D_0\phi}{q_0 R} \frac{1}{s^{3/2}} \frac{\partial s}{\partial \bar{x}} =\bar{C}_1 \frac{R}{\delta_R}\frac{S_e}{S} \sqrt{\frac{\bar{N}_c}{\phi}}
\end{equation}
If the flux density is known in a steady state, the coefficient of diffusion can be obtained by fitting the observed profiles of $s(\bar{x})$ to get $\bar{C}_1$. So that the ratio of the area of entrances and exits $S_e/S$, the main connectivity parameter, is parametrized by the non-dimensional parameters $\delta_R/R$ and $\phi$, since $\bar{N}_c=\bar{N}_c(\phi)$.

\begin{table*}[t]
\resizebox{1.1\columnwidth}{!}{
    \begin{tabular}{ | c | c | c | c | c | c | }
      \hline 
      Number of nodes $n_T$ & $\alpha_{mn}$  &  $\phi$  & $\bar{N}_c$  & $\bar{q}_s$ & $S_e/S$  \\  
      \hline

6400 & $1$  &  $0.5$  & $0.64$   & $-0.15\pm 0.008$ & $0.49\frac{\delta_R}{R}$  \\  
      \hline
			
			6400 & Random  &  $0.5$  & $0.64$   & $-0.056\pm 0.005$ & $0.17\frac{\delta_R}{R}$  \\  
      \hline
			
			2300 & $1$  &  $0.7$  & $0.23$  & $-0.038\pm 0.004$ & $0.25\frac{\delta_R}{R}$  \\  
      \hline
			
			2300 & Random  &  $0.7$  & $0.23$ & $-0.015\pm 0.001$ & $0.1\frac{\delta_R}{R}$  \\  
      \hline
                          
      \hline                    
    \end{tabular} }
    \caption{Simulation results in the steady state of the quasi-one-dimensional microscopic network model with triangular tessellation (three neighbours per a node) at different node densities, total number of nodes $n_T$. The saturation levels at the interval ends are fixed at $s_1=0.1\%$ and $s_2=20\%$. All data has been averaged over five independent simulations.}
    \vspace{5pt}
    \label{Table1}
\end{table*}

\begin{table*}[t]
\resizebox{1.1\columnwidth}{!}{
    \begin{tabular}{ | c | c | c | c | c | c |  }
      \hline 
      Number of nodes $n_T$ & $\alpha_{mn}$  &  $\phi$  & $\bar{N}_c$  & $\bar{q}_s$ & $S_e/S$  \\  
      \hline

6400 & $1$  &  $0.5$  & $0.64$  & $-0.27\pm 0.01$ & $0.88\frac{\delta_R}{R}$  \\  
      \hline
			
			6400 & Random  &  $0.5$  & $0.64$  & $-0.11\pm 0.005$ & $0.36\frac{\delta_R}{R}$  \\  
      \hline
			
			2300 & $1$  &  $0.7$  & $0.23$  & $-0.075\pm 0.004$ & $0.49\frac{\delta_R}{R}$  \\  
      \hline
			
						2300 & Random  &  $0.7$  & $0.23$ & $-0.03\pm 0.002$ & $0.19\frac{\delta_R}{R}$  \\  
      \hline
                          
      \hline                    
    \end{tabular} }
    \caption{Simulation results in the steady state of the quasi-one-dimensional microscopic network model with quadrilateral tessellation (four neighbours per a node) at different node densities, total number of nodes $n_T$. The saturation levels at the interval ends are fixed at $s_1=0.1\%$ and $s_2=20\%$.  All data has been averaged over five independent simulations.}
    \vspace{5pt}
    \label{Table2}
\end{table*}

\subsubsection*{Steady state distributions and the network connectivity factor}

In the simulations, the quasi one-dimensional network setup corresponding to the macroscopic model (\ref{S-profile}) evolved in time till the flux density arrived at a uniform distribution in the bulk within the tolerance of $5-10\%$. We used different total number of points in the fixed simulation domain (side size $\bar{X}=100$) $n_T$ and two types of networks, with three neighbours (triangular tessellation) and four neighbours (quadrilateral tessellation) per each node,  namely, $n_T=2300$, $\bar{N}_c=0.23$ and $\phi=0.7$, and $n_T=6400$ , $\bar{N}_c=0.64$ and $\phi=0.5$. The boundary values of saturation have been set to $s_1=0.1\%$ and to $s_2=20\%$ to cover the whole range, where the super-fast regime may be expected. We have also used two different models for the link permeability parametrised by the non-dimensional coefficients $\alpha_{ij}$, when either all $\alpha_{ij}=1$ or they were randomly, but uniformly distributed in the interval $0\le \alpha_{ij} \le 1$, such that the average $<\alpha_{ij}>=1/2$.   

\begin{figure}[ht!]
\begin{center}
\includegraphics[trim=0.5cm 2.cm 0.5cm 0cm,width=\columnwidth]{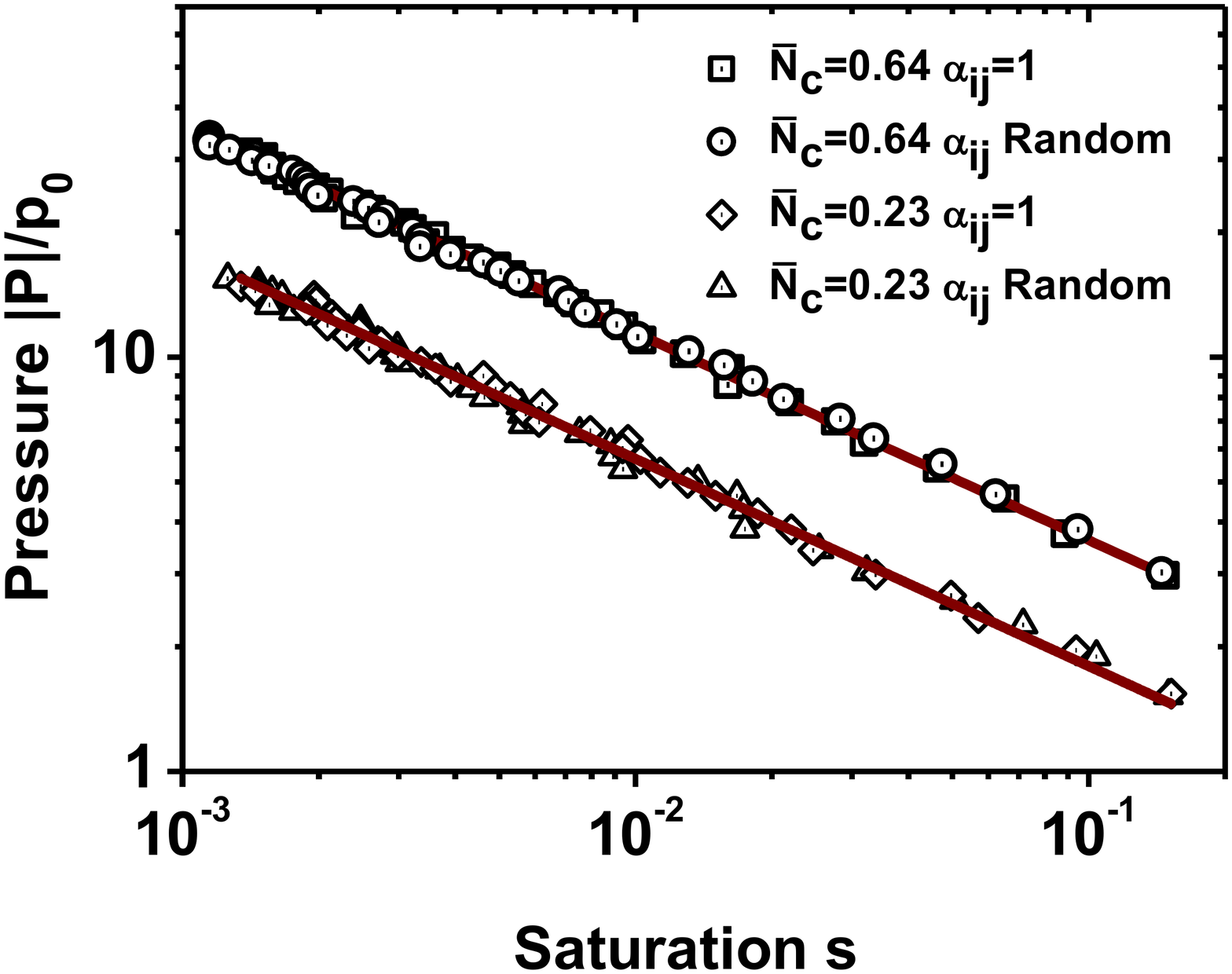}
\end{center}
\caption{Averaged, reduced capillary pressure $|P|/p_0$ as a function of saturation $s$ in the model set-up with three nodal neighbours at $s_1=0.001$ and $s_2=0.2$, and at different values of $\bar{N}_c$, $\phi=0.5$, $\bar{N}_c= 0.64$ and $\phi=0.7$ and $\bar{N}_c= 0.23$, and different distributions of $\alpha_{ij}$. The numerical data are shown by symbols and the solid lines (brown) indicate the fitting function $\frac{|P|}{p_0}=A_f s^{-1/2}$, $A_f=\sqrt{\frac{\bar{N}_c}{\phi}}$.} 
\label{Fig3}
\end{figure}

What do we observe in simulations with the microscopic model? After reaching a steady state, when the flux density is constant in the flow domain, the distribution of pressure as a function of saturation, Fig. \ref{Fig3}, was found to be in very good agreement with that anticipated in the macroscopic model (\ref{Pressure-saturation}), which is in a non-dimensional form
\begin{equation}
\label{P-S-NDF}
\bar{P}=-\sqrt{\frac{\bar{N}_c}{\phi}} \frac{1}{\sqrt{s}}.
\end{equation}

\begin{figure}[ht!]
\begin{center}
\includegraphics[trim=0.5cm 2.cm 0.5cm 0cm,width=\columnwidth]{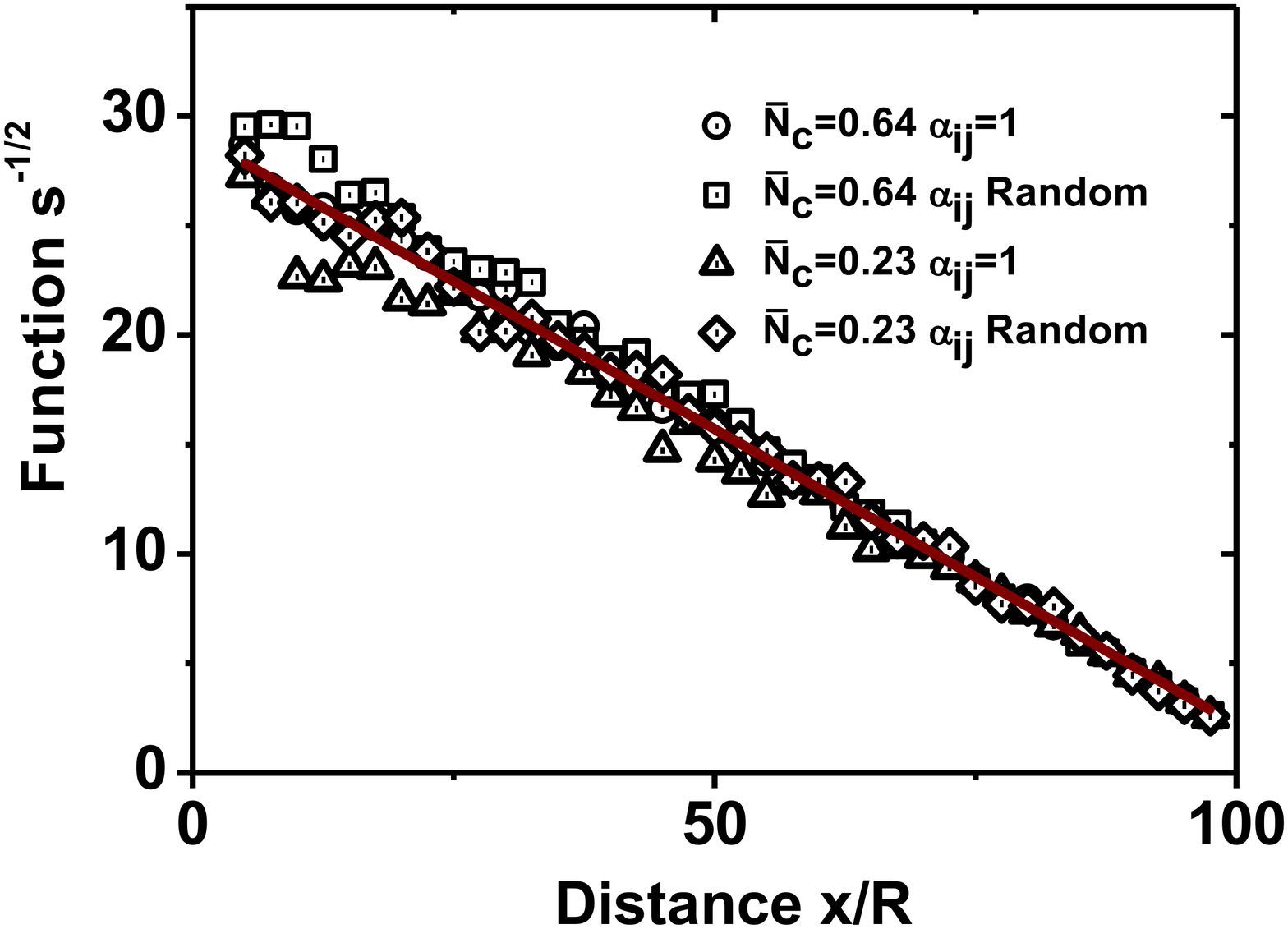}
\end{center}
\caption{Average saturation $s^{-1/2}$ as a function of the reduced distance $\bar{x}=x/R$ in the model set-up with three nodal neighbours at $s_1=0.001$ and $s_2=0.2$, and at different values of $\bar{N}_c$, $\phi=0.5$, $\bar{N}_c= 0.64$ and $\phi=0.7$ and $\bar{N}_c= 0.23$, and different distributions of $\alpha_{ij}$. The numerical data are shown by symbols and the solid lines (brown) indicate the fitting function $s^{-1/2}=A_s+B_s \bar{x}$ at $A_s=29.2\pm 0.2$ and $B_s=-0.27\pm 0.004$.} 
\label{Fig4}
\end{figure}

As one can observe, Fig. \ref{Fig4}, the saturation profiles $s(\bar{x})$ are in accord with those anticipated from the macroscopic model (\ref{S-profile}). One can conclude that on average the behaviour of the network model can be adequately described by the macroscopic equations.   

The results of simulations involving network models with different parameters are summarized in Tables \ref{Table1} and \ref{Table2}. The connectivity factor $S_e/S$ obtained in the simulations strongly depends (non-linearly) on the assumptions made about the conductivity of the links $\alpha_{mn}$ and, of course, on the node density $\bar{N}_c$, that is on the porosity $\phi$. In general, the lower the porosity, the larger the conductivity, since more links are available to transfer the liquid. 

The non-trivial behaviour is observed when at a fixed value of $\bar{N}_c$, the conductivity of the links becomes a random distribution. One can see from the tables, that while the mean value of $<\alpha_{mn}>=0.5$, the connectivity factor $S_e/S$ changes almost three times. Natural paper materials have rather random structures on the microscopic level, so that such changes should be taken into account. The result also implies that a small number of impurities obstructing the capillary flow may substantially reduce permeability of textured materials, as the super-fast diffusion mechanism is particularly sensitive to the tortuosity of the pathways. 

At the same time, the scaling factor of about $2$, which is expected to occur in different microscopic connectivity models, that is when changing from the triangular tessellation (three neighbours per a node) to the quadrilateral one (four neighbours per a node), is clearly observed in the average flux density values, Tables \ref{Table1} and \ref{Table2}. 

\section{Observation of liquid spreading in paper porous materials}

\subsection{Experimental procedures}

The experiments were designed to study interaction of a single liquid drop with a porous matrix, its subsequent penetration and spreading in the porous material within the framework of the witness card technique, which is widely used for accurate determination of the particle size distributions to assess the effectiveness of spraying in applications~\cite{WCT1991}. The emphasis in the current research was on the analysis of the general trends of liquid spreading at low saturation levels. 

Our previous study of liquid dispersion in particulate porous media has shown that evolution of the wetting front in the later stages of the spreading, when the saturation level is below a critical value $s_c\approx 10\%$, follows a universal power law, when the wetting spot diameter $D(t)$ as a function of time $t$ obeys $D(t)\propto t^{\beta}$, where the time is measured from the onset of the low saturation regime and the exponent $\beta=1/(N_d+1)$ is a function of the  dimension $N_d$ of the spreading domain only~\cite{Lukyanov2019}. In particular, in our case, the spreading geometry in papers is two-dimensional, $N_d=2$, so that it is anticipated that $\beta=1/3$. 

We note here that the power law dependencies have been observed in both the experiments in particulate porous media and the numerical and asymptotic analysis of the superfast diffusion model. Therefore, the power law dependence with the exponent $\beta=1/(N_d+1)$ is very characteristic for the low saturation regime of spreading, when the non-linear dynamics is governed by the so-called super-fast non-linear diffusion mechanism, details can be found in~\cite{Lukyanov2019}. Consequently, observation of the wetting spot evolution can reveal, in principle, the character of the diffusion process, and indicate that the diffusion process at low saturation levels in fibrous materials is also driven by the super-fast diffusion mechanisms, as that in the particulate porous media.  

Note, the exponent is a very good indicator. One can easily distinguish between the power law expected in fully saturated porous matrices and that in the case of the super-fast diffusion at low saturation values. For example, $\beta=1/2$ in two-dimensional fully saturated cases~\cite{Stone2012}.

On the other hand, as previous studies of liquid spreading in particulate porous media had clearly demonstrated, any detailed quantitative characterisation of the dispersion process at low saturation levels requires very detailed information about the porous media structure~\cite{Penpark2018, Penpark2019, Lukyanov2019}. Therefore, there was no any detailed characterisation of the paper material itself in the current study, such as their micro-structure. This will be the subject of future research, which may require a completely different approach.

\begin{figure}[ht!]
\begin{center}
\includegraphics[trim=0.5cm 3.cm 0.5cm 0cm,width=\columnwidth]{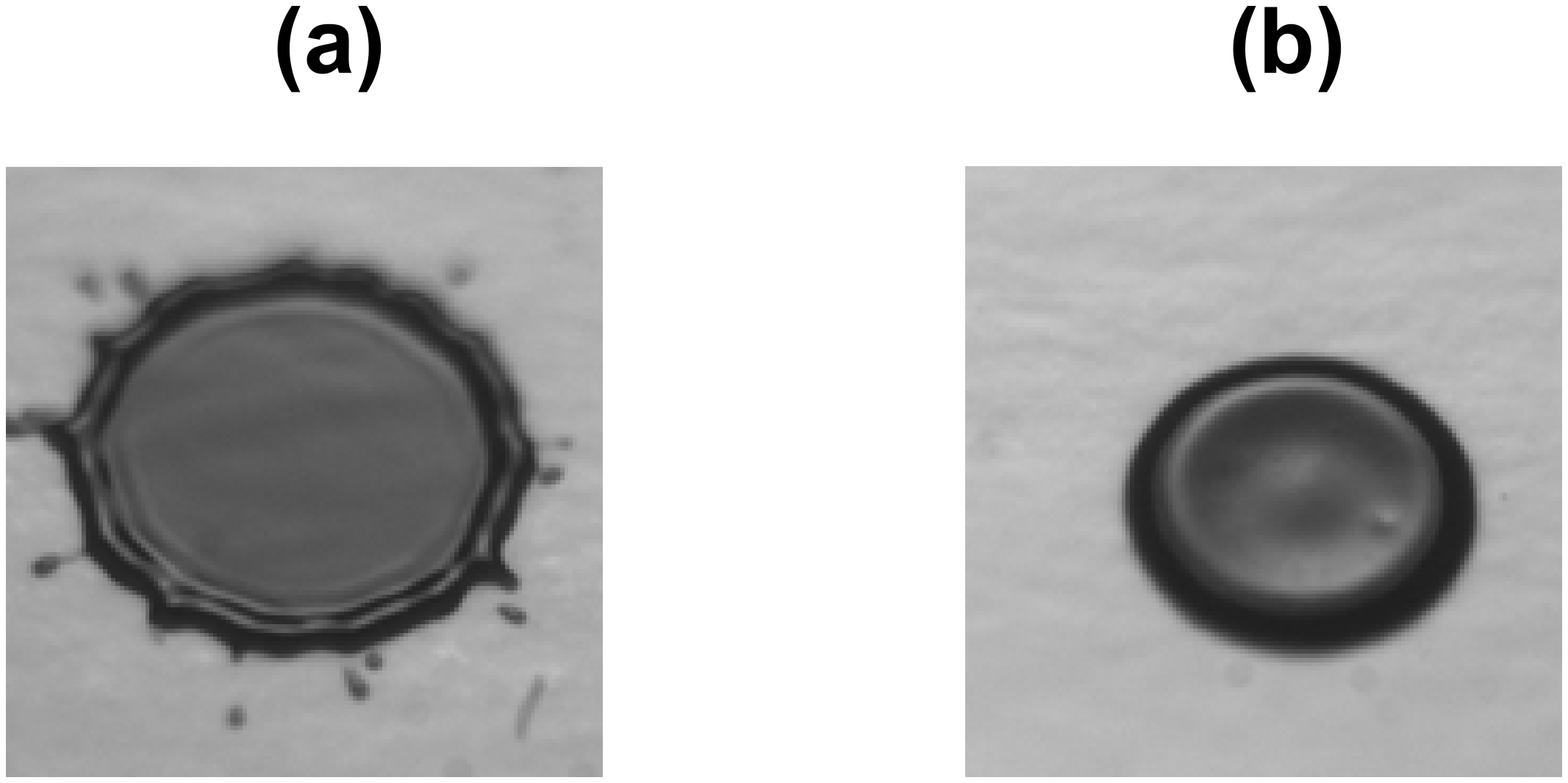}
\end{center}
\caption{Drop impact at $t=1.6\,\mbox{ms}$ after the initial contact at the impact velocity $3.1\,\mbox{m/s}$: (a) neat TBP (b) TBP with 3.8\% of PSBMA.} 
\label{Fig7}
\end{figure}

In the experiments, a single liquid drop of a controlled volume has been dispatched from the drop generator whose position was adjustable. After separation from the generator head, the drop was accelerated by gravity up to its terminal velocity in the air. The variable positioning of the drop generator allowed for easy control of the drop impact velocity. The process of the falling drop splashing and spreading over the substrate was recorded by a high-speed video camera with the frame rate up to $20000 \, \mbox{fps}$ and a spatial resolution $50\,\mbox{pixel/mm}$. The recording was synchronized with a system of drop detection, which also made it possible to accurately measure the velocity at the time of the impact, Fig. \ref{Fig7}. In the current study, we used two characteristic values of the impact velocity $u=0.2\, \mbox{m/s}$ and $u=3.1\, \mbox{m/s}$.

The test liquid was neat tributyl phosphate (TBP, molar weight $266.32$ g/mol), a low-volatility organophosphate compound, dyed with Calco red or blue oil ($0.11\%-0.5\%$ mass concentration respectively). To introduce non-Newtonian effects and variations of viscosity, the neat (dyed) TBP solution was mixed with $3.8 \%$ (mass concentration) of Poly(Styrene-Butyl Methacrylate) (PSBMA).

\bigskip

The neat TBP solution has liquid viscosity $\mu=3.88\,\mbox{mPa}\cdot\mbox{s}$ and surface tension $\gamma=28\pm 1\, \mbox{mN}/\mbox{m}$ measured in our laboratory at $20^{\circ}\, \mbox{C}$. The addition of the polymer into the pure TBP liquid resulted in substantial increase in the liquid viscosity $\mu_{P}\approx 340\,\mbox{mPa}\cdot\mbox{s}$ at practically identical values of the surface tension $\gamma\approx 27\, \mbox{mN}/\mbox{m}$, basically introducing non-Newtonian behaviour during the first, short lasting stage of the impact to avoid formation of satellite droplets, the so-called corona of the splashing droplet, Fig. \ref{Fig7}. The details of the properties of the polymer solution, also used in viscoelastic aerobreakup studies, can be found in ~\cite{Theo2013}. 

\bigskip
\begin{figure}[ht!]
\begin{center}
\includegraphics[trim=0.5cm 1.cm 0.5cm 0cm,width=\columnwidth]{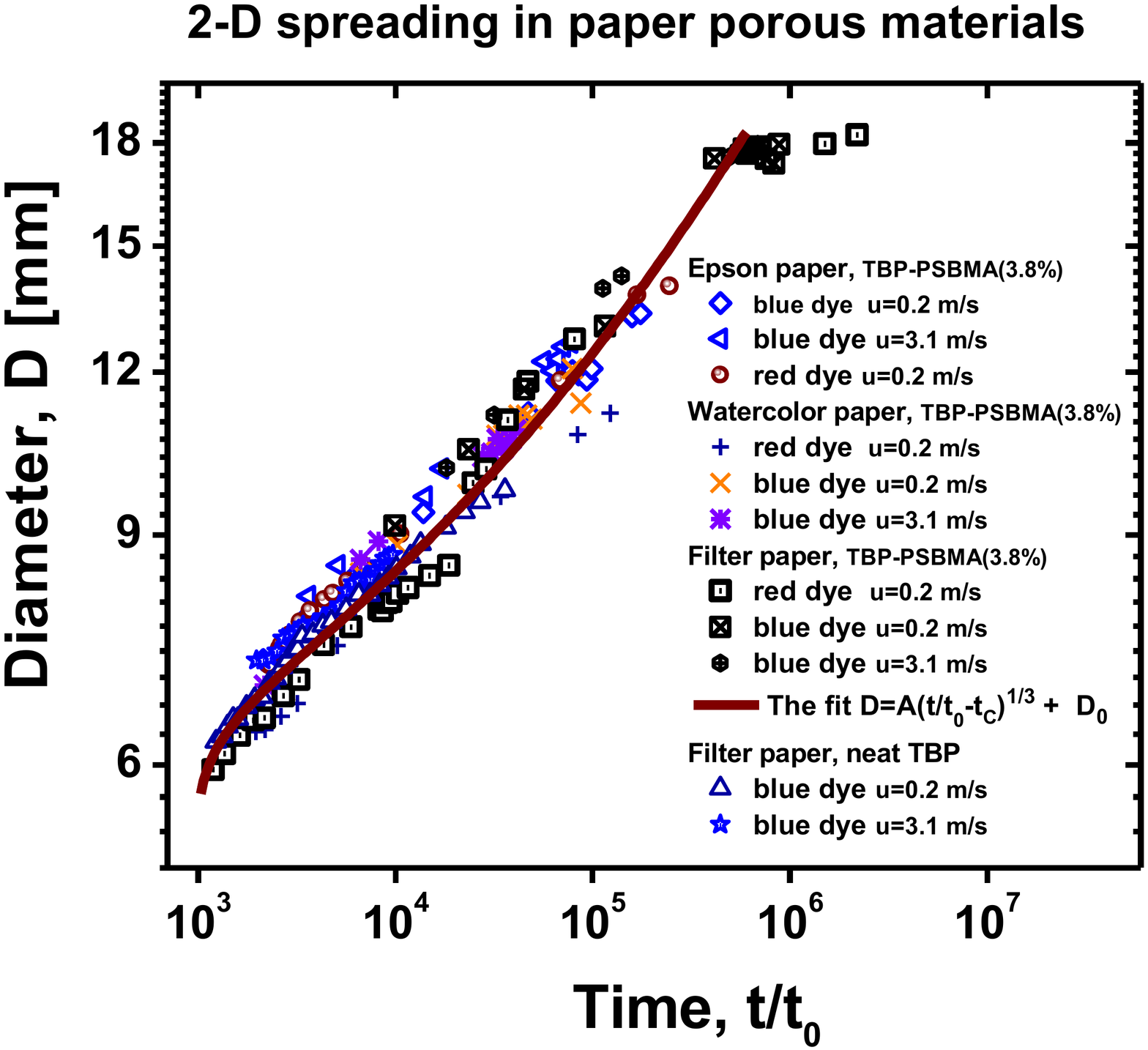}
\end{center}
\caption{The spot diameter $D$ as a function of the reduced time $t/t_0$. Experimental data are shown by symbols at different impact velocities ($u=0.2\, \mbox{m/s}$ and $u=3.1\, \mbox{m/s}$), for different liquids (neat TBP and TBP with 3.8\% PSBMA), different dyes (the Calco blue and red oils at $0.5\%$ and $0.11\%$ mass concentrations respectively) and different paper matrices (Epson, Filter and Watercolor papers). The solid line (brown) is the power-law fit $D=A(t/t_0-t_C)^{1/3}+D_0$ at $A=0.16\,\mbox{mm}$, $t_C=10^3$ and $D_0=5.2\,\mbox{mm}$. The data collapsed into a master curve by using the characteristic time scale: for the Filter paper and TBP+PSBMA $t_0=1/9\,\mbox{s}$, the Watercolor paper and TBP+PSBMA $t_0=2\,\mbox{s}$, for the Epson paper and TBP+PSBMA $t_0=1\,\mbox{s}$ and for the Filter paper and neat TBP $t_0=1/900\,\mbox{s}$.} 
\label{Fig8}
\end{figure}

\subsection{Results and discussions}
As we have already discussed,  the detailed description of the liquid dispersion process in the papers, which requires microscopic information on the porous paper matrix, will be the subject of future research, so that here, we only analyse the general trend by observing the wetting spot diameter $D(t)$ as a function of time. Our prime concern is the long-time evolution of the spot diameter, which is shown in Fig. \ref{Fig8}. 

In the experiments, all drops were of a fixed volume of $2.2\, \mu L$. There were three different fibrous substrates: Epson paper ($80 \mbox{g}/\mbox{m}^2$), Watercolour paper ($300 \mbox{g}/\mbox{m}^2$) and Filter paper ($70 \mbox{g}/\mbox{m}^2$). It appears, though not surprisingly, that the long-time evolution of the wetting spot diameter (after some initial relaxation time $t_C$, that is at $t>t_C$) on all samples in different conditions can be effectively reduced to a single master curve by renormalizing time $t/t_0$, where the characteristic time $t_0$ only depends on the liquid viscosity and the substrate material, Fig. \ref{Fig8}, but, essentially, is independent of the initial conditions, such as the impact velocity, and the visualization materials (red or/and blue oils).

As one can clearly see from the figure, the spreading law $D(t/t_0)\propto (t/t_0 - t_C)^{1/3}$ is well observed at $t/t_0\ge t_C$ indicating that indeed the process of spreading after some initiation time $t_C$ follows the super-fast non-linear diffusion model. As one can also observe, the long-time evolution characteristic behaviour (the exponent $\beta$) is insensitive to the impact drop velocity, the type and concentration of the visualization liquid (red or blue Calco dyes), the substrate and liquid properties despite obvious difference in the initial conditions and the substrate and liquid properties, Fig \ref{Fig7}. One may also note that in spite of the non-Newtonian character of the polymer solution, the observed effect while switching from the neat TBP to its $3.8\%$ polymer solution is simply down to the change in the zero shear rate viscosity from $\mu=3.88\,\mbox{mPa}\cdot\mbox{s}$ to $\mu_{P}\approx 340\,\mbox{mPa}\cdot\mbox{s}$, corresponding to the change in $t_0$ from $t_0=1/900\,\mbox{s}$ to $t_0=1/8\,\mbox{s}$, assuming $t_0\propto \mu$. 

\section*{Conclusions}

In conclusion, the diffusion process at low saturation levels in fibrous porous materials is shown to be fully compatible with that anticipated from the macroscopic super-fast diffusion model. The long-time behaviour is well consistent with the model predictions, but further work is required to link microscopic parameters of the fibrous porous matrix with the macroscopic parameters of the evolution to enhance the predictive power of the model. What's important for applications is that the long-time behaviour is insensitive to the initial conditions (impact velocity and the character of the initial splash), but only depends on the liquid properties (viscosity) and the properties of the substrate through a single parameter $t_0$. It is important, that the character of the evolution law is universal, such that the exponent $\beta=1/3$, and it can only be influenced by the geometry of the diffusion domain, its dimension $N_d$, $\beta=1/(N_d+1)$.

\end{document}